\documentclass[twocolumn]{aastex631}

\usepackage{amsmath, color, amssymb, latexsym, graphicx, lipsum, url}
\usepackage{bm}
\newcommand{\bfu}{\bm{u}}

\newcommand{\bfb}{\bm{b}}

\newcommand{\bfk}{\bm{k}}
\newcommand{\bnabla}{\bm{\nabla}}

\newcommand{\cs}{c_s}
\newcommand{\M}{\mathcal{M}}
\newcommand{\Ma}{\mathcal{M}_{\rm A}}

\newcommand{\Rm}{\mathrm{Rm}}
\renewcommand{\Re}{\mathrm{Re}}

\newcommand{\V}{\mathcal{V}}
\newcommand{\Exp}[2]{\left\langle{#1}\right\rangle_{#2}}
\renewcommand{\d}[1]{\ensuremath{\operatorname{d}\!{#1}}}

\newcommand{\Mt}{\mathcal{M}_t}

\shorttitle{Compressibility in high-beta plasma}
\shortauthors{Bandyopadhyay, Beattie, \& Bhattacharjee}

\begin{document}

\correspondingauthor{\\
$^{\dagger}$Riddhi Bandyopadhyay: \href{mailto:riddhib@princeton.edu}{riddhib@princeton.edu}\\
$^{\ddagger}$James R. Beattie: \href{mailto:james.beattie@princeton.edu}{james.beattie@princeton.edu}\\
These authors contributed equally to this work.
}

\title{Density fluctuation - Mach number scaling in compressible, high plasma beta turbulence: \\ \textit{in-situ} space observations and high-Reynolds number simulations}

\author[0000-0002-6962-0959]{Riddhi Bandyopadhyay$^{\dagger}$}
\affiliation{Department of Astrophysical Sciences, Princeton, NJ 08544, USA}

\author[0000-0001-9199-7771]{James R. Beattie$^{\ddagger}$}
\affiliation{Department of Astrophysical Sciences, Princeton, NJ 08544, USA}
\affiliation{Canadian Institute for Theoretical Astrophysics, University of Toronto, Toronto, M5S3H8, ON, Canada}

\author[0000-0001-6411-0178]{Amitava Bhattacharjee}
\affiliation{Department of Astrophysical Sciences, Princeton, NJ 08544, USA}

\begin{abstract}
    Understanding the nature of compressible fluctuations in a broad range of turbulent plasmas, from the intracluster medium to the solar wind, has been an active field of research in the past decades. Theoretical frameworks for weakly compressible MHD turbulence in an inhomogeneous background magnetic field predict a linear scaling of the normalized mass density fluctuation ($\delta \rho / \rho_0$), as a function of the turbulent Mach number ($\Mt$), $\delta \rho / \rho_0 \propto \Mt$. However, so far the scaling relation has been tested only using moderate to low plasma beta ($\beta \lesssim 1$) solar wind observational data where the compressibility is weak $\delta \rho / \rho_0 \sim 0.1$. Here, we combine NASA’s Magnetospheric Multiscale Mission data in Earth's magnetosheath, where $\beta \sim 10$ is high, and $\beta \sim 1$ highly-compressible magnetohydrodynamic turbulence simulations at unprecedented resolutions. Both show that $\delta \rho / \rho_0 \propto \Mt$ holds across a broad range of $\delta \rho / \rho_0$, $\Mt$ and $\beta$, demonstrating that $\delta \rho / \rho_0 \propto \Mt$ is a robust compressible turbulence relation, going beyond the asymptotics of the weakly compressible theory. We discuss the findings in the context of understanding the nature of strongly compressible turbulent fluctuations and the driving parameter in astrophysical and space plasmas.
\end{abstract}
\keywords{Plasma Turbulence --- Astrophysical Plasmas -- Solar wind}

\section{Introduction}\label{sec:intro}
    Most of the visible Universe is made of ionized plasmas, and most naturally occurring plasmas exist in a turbulent state~\citep{Matthaeus2011SSR}, ranging from the interstellar medium~\citep{Elmgreen2004_ISM_turbulence,Brandenburg2013SSR_review,Beattie2024_10k}, intracluster medium~\citep{Mohapatra2019_ICM_turbulence,Kunz2022_ICM_physics_review} to solar wind and planetary magnetospheres \citep{Verscharen2019LRSP}. Turbulent fluctuations in plasmas have important effects on heating, transport, and overall evolution and structure in these systems~\citep{Usmanov2014ApJ, Adhikari2023ApJ_high-beta}. An important parameter controlling the dynamics in turbulent plasmas is the plasma beta $\beta$, defined as 
    \begin{equation} \label{eqn:beta}
        \beta = \frac{n \,k_B\,T}{B^2/\left(2\,\mu_0\right)},
    \end{equation}
    where $n$ is the plasma number density, $k_B$ the Boltzmann constant, $T$ is the temperature, $B$ is the magnetic field, and $\mu_0$ the vacuum permeability.  Physically, $\beta$ is the ratio of the thermal pressure, to the magnetic pressure $P_{\rm mag}$. Although the interplanetary solar wind is mostly characterized by $\beta \sim 1$, high-beta plasmas $(\beta \gg 1)$ are prevalent in many astrophysical environments, e.g., in the warmer phases of the interstellar medium \citep{Ferriere2020_reynolds_numbers_for_ism} and the intracluster medium \citep{Kunz2022_ICM_physics_review}, where $\beta \gg 1$ plays an essential role in facilitating the conditions for mirror and firehose instabilities. 

    For an isothermal plasma, $\beta = 2c_s^2 / v_A^2$ where $c_s$ is the sound speed, and $v_A$ the Alfv\'en speed. Hence, in some ways, $\beta$, which controls the speed of the fast wave relative to the Alfv\'en speed, characterizing how compressible a plasma is. The properties and origins of compressible fluctuations in turbulent space plasmas are still not well understood~\citep{Klainerman1981CPAM_singular, Klainerman1982CPAM_compressible, Shebalin1988JPP_density, Du2023ApJ_density}. Density fluctuation $(\delta \rho = \left\langle (\rho - \left\langle \rho \right\rangle)^2\right\rangle^{1/2})$ in the solar wind is typically small compared to background density $(\rho_0 = \langle \rho \rangle)$. Consequently, theoretical frameworks~\citep{Matthaeus1988PoF_NIMHD, Matthaeus1991JGR_NIMHD, Zank1993PoF_NIMHD} have been developed for small-amplitude compressive fluctuations $(\delta \rho / \rho_0 \ll 1)$, assuming a small turbulent Mach number, which is defined
    \begin{eqnarray}\label{eq:mach}
    \Mt = \frac{\left\langle (u - \left\langle u \right\rangle)^2\right\rangle^{1/2}}{\left\langle c_s \right\rangle} = \frac{\delta u}{c_s},  
    \end{eqnarray}
    with $\delta u$ as the rest-frame rms velocity fluctuation and $c_s$ as the mean sound speed. For a homogeneous background field, the nearly-incompressible MHD (NI-MHD) theories predict that the normalized density fluctuations scale with the square of $\Mt$, 
    \begin{eqnarray}\label{eq:m2}
        \delta \rho / \rho_0 \propto \Mt^2. 
    \end{eqnarray}
    However, a further generalization of NI-MHD theory, referred to here as the weakly compressible MHD theory for the case of an inhomogeneous background field~\citep{Bhattacharjee_1998_weakly_compressible_solar_wind, Bhattacharjee1999JGR_four-field, Hunana2010ApJ_inhomogeneous} predicts a linear relationship of normalized density fluctuations scaling with $\Mt$, 
    \begin{eqnarray}
        \delta \rho / \rho_0 \propto \Mt\label{eq:m1}. 
    \end{eqnarray}
    For supersonic turbulence, $\Mt \gg 1$, ubiquitous in the star-forming, colder phases of the interstellar medium \citep{Krumholz2015_star_formation_text}, with large magnetic field \citep{Beattie2020c} and mass density inhomogeneities \citep{Federrath2013_universality,Beattie2020}, a further model has been proposed, based on the Rankine-Hugoniot shock-jump conditions. For isothermal, hydrodynamic shocks \citet{Padoan1997_imf} and \citet{Passot1998} showed that the density variance can be related to $\Mt$ by
    \begin{align} \label{eq:linear_mach}
        \delta \rho / \rho_0 = b \Mt,
    \end{align}
    where $b$ is the so-called `driving parameter,' which depends nonlinearly on how the turbulence is driven, i.e., if energy is injected with compressible modes ($b=1$ for purely $|\bnabla\times\bfu| = 0$ modes) or incompressible modes ($b=1/3$ for purely $\bnabla\cdot\bfu = 0$ modes) \citep{Federrath2010_solendoidal_versus_compressive}. This was later generalized to high-$\beta$ shocks in \citet{Molina2012_dens_var}, non-isothermal shocks in \citet{Nolan2015} and low-$\beta$ shocks in \citet{Beattie2021_multishock}.

    Numerous studies have been conducted to verify the relation of density fluctuation and turbulent Mach number, using both observational space and astrophysical data~\citep{Matthaeus1991JGR_NIMHD, Tu1994JGR_compressive, Bavassano1995JGR_Density_fluctuations, Federrath2016_brick,Menon2020,Menon2020b, Sharda2021_driving_mode,Gerrard2023_driving_parameter_LMC,Cuesta2023ApJL_compressible, Zhao2025ApJL_Transonic} and numerical simulations~\citep{Padoan1997_imf,Passot1998,Kowal2007ApJ_density, Federrath2010_solendoidal_versus_compressive,Price2011,Burkhart2012,Nolan2015,Pan2019,Mohapatra2020b, Beattie2021_multishock,Dhawalikar2022_shock_driven_turbulence_sim,Du2023ApJ_density}, and mostly support the linear relation (\autoref{eq:m1} and \autoref{eq:linear_mach}). Observationally in space plasmas, to the best of our knowledge, the $\delta\rho-\Mt$ scaling relations have not been tested for high-$\beta$ regime, and nor have they been tested on a scale-by-scale basis in simulations, far away from the modes that drive the turbulence. Our goal in this work is to then further examine the scaling relation between $\delta\rho-\Mt$ for high-beta plasma. To test this, we utilize \textit{in-situ} data in Earth's magnetosheath, where the effect of spatial inhomogeneities is strong and the typical $\beta \sim 10$. Along with the \textit{in-situ} data, we make use of an unprecedentedly high-resolution MHD simulation at Reynolds number $\rm{Re} \gtrsim 10^6$ and $\beta \sim 1$. Regardless of $\beta$, within $1\sigma$ uncertainties, both data support a linear scaling $\delta\rho/\rho_0 \propto \Mt$, even sharing (statistically) the same proportionality constants. This potentially demonstrates the importance of inhomogeneous fluctuations in compressible, turbulent plasmas, and supports the weakly compressible MHD turbulence theory in \citet{Bhattacharjee_1998_weakly_compressible_solar_wind}.

    This paper is organized as follows. In \autoref{sec:methods} we discuss the observational and simulation data that we use in this study. In \autoref{sec:results} we calculate scale-dependent $\delta\rho/\rho_0$ and $\Mt$ curves from the simulation and combine them with the observational data to reveal that $\delta\rho/\rho_0 \propto \Mt$ over a broad range of $\delta\rho/\rho_0$ and $\Mt$, regardless of $\beta$. Finally, in \autoref{sec:disc} we discuss the implications our results have for weakly compressible MHD theory and space and astrophysical plasmas, and in particular driving parameter measurements that may be contaminated by cascade effects.

    \begin{figure*}
       \centering
      \includegraphics[width = 0.95\linewidth]{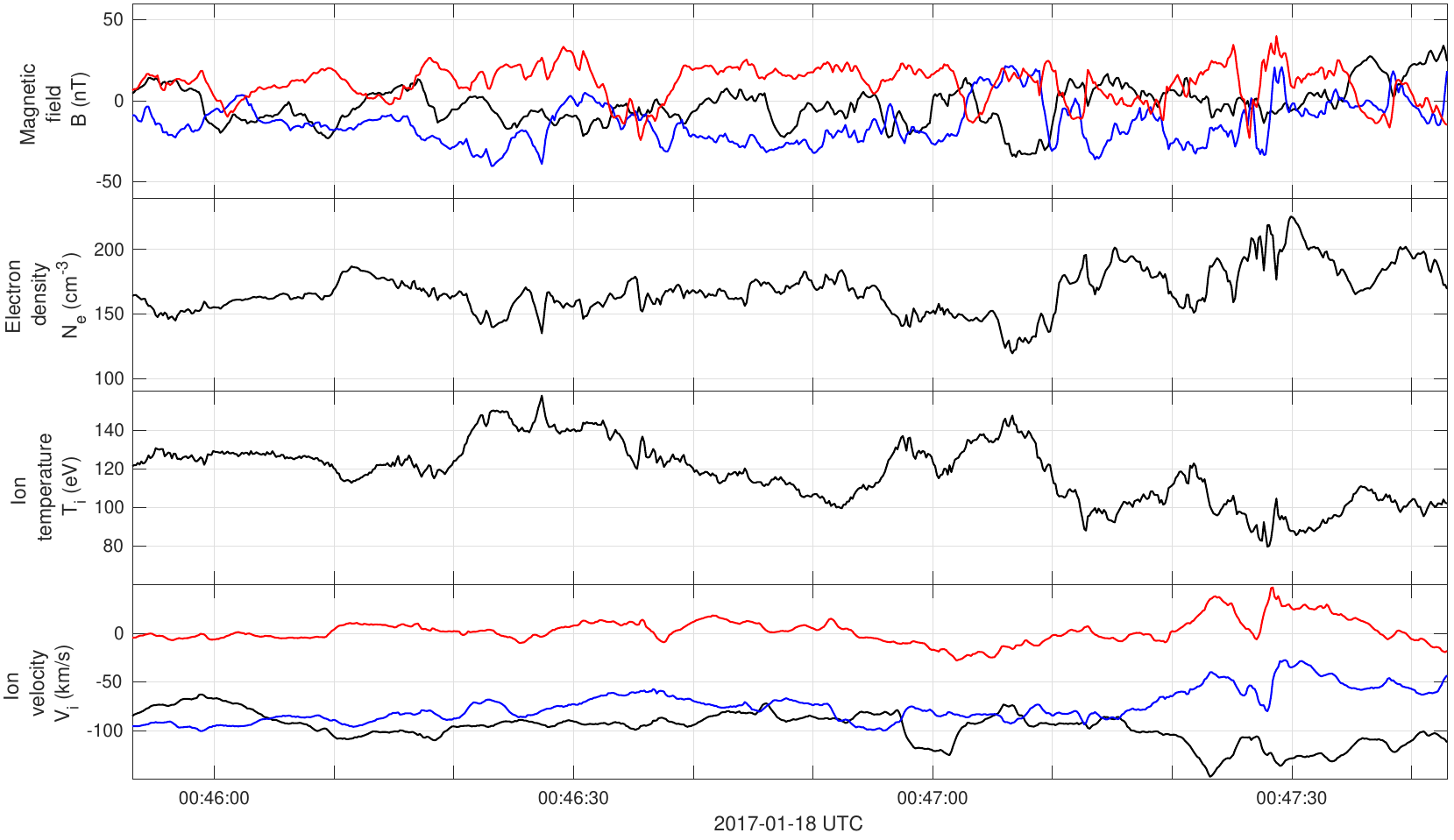}
       \caption{An example of the MMS data in Earth's turbulent magnetosheath. The data shown are from the FGM and FPI instruments on-board the MMS1 spacecraft. The top panel shows the magnetic field measurements in GSE coordinates; the second panel shows the electron density; third panel shows the ion temperature; and the bottom panel shows the ion velocity in GSE coordinates. Each panel shows significant temporal turbulent fluctuations, which, under the assumption of Taylor's frozen-in hypothesis, we can equate to sampling the spatial turbulent fluctuations.}\label{fig:mms}
    \end{figure*}

    \begin{figure*}
       \centering
      \includegraphics[width = \linewidth]{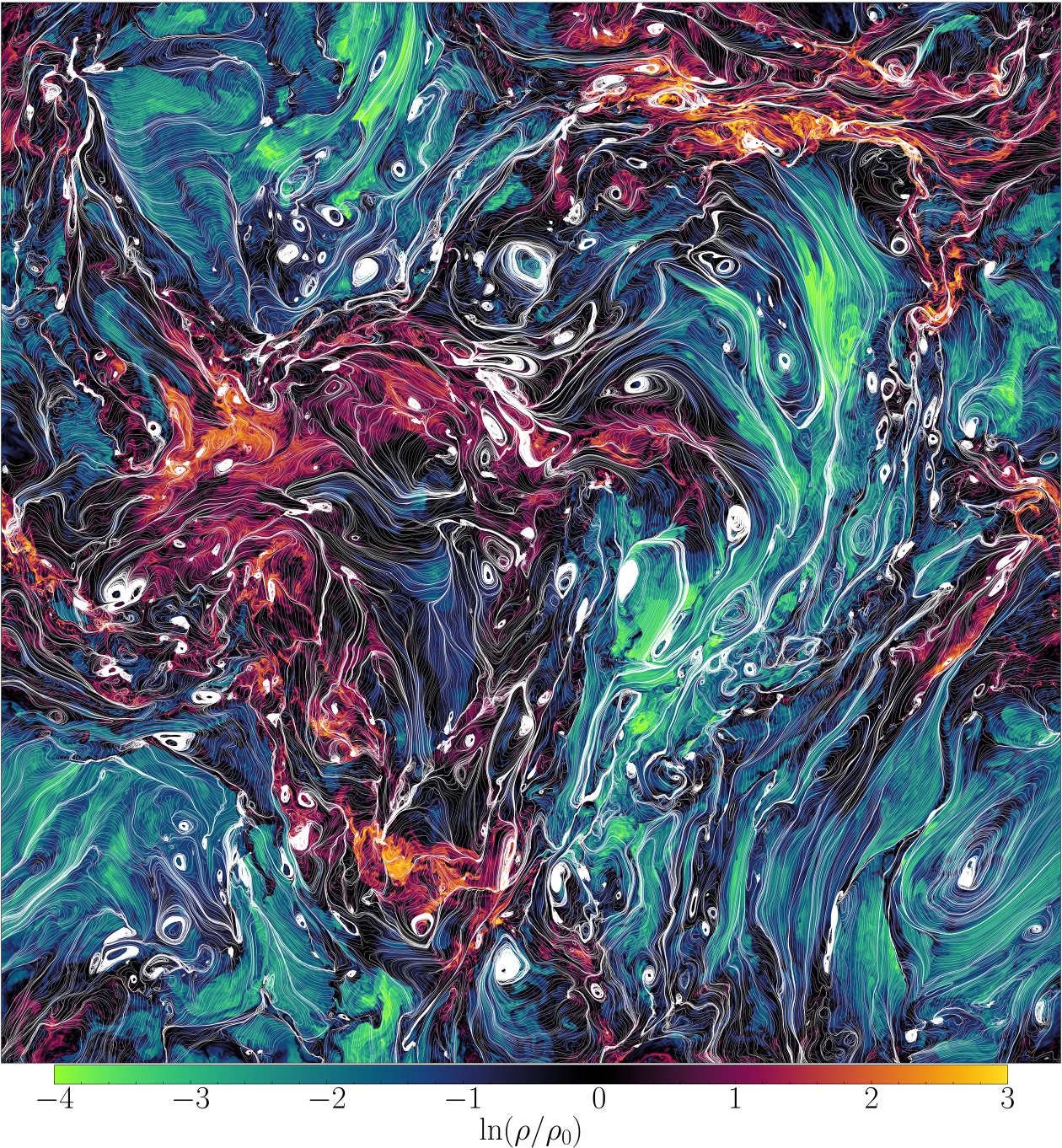}
       \caption{A two-dimensional slice of the logarithmic mass density fluctuations, $\ln(\rho/\rho_0)$, where $\rho_0$ is the volume-average, overlaid with in-plane magnetic field streamlines shown in white. The mean density $\rho = \rho_0$ is shown in black, over-densities $\rho > \rho_0$ in yellow and under-densities $\rho < \rho_0$ in green. Due to the strong $\rho/\rho_0$ contrasts and the coherent, over-dense filamentary structures and deep under-dense voids the mass-density (and magnetic field) is highly-inhomogeneous. More details of the $10,\!080^3$ simulation are shown in \citet{Beattie2024_10k}.}
       \label{fig:dens}
    \end{figure*}
 
\section{Methods}\label{sec:methods}
\subsection{MMS Observations}
    We use \textit{in-situ} data collected by NASA’s Magnetospheric Multiscale (MMS) mission~\citep{Burch2016SSR, Burch2016Science} in Earth's magnetosheath. The magnetosheath consists of the shocked solar wind plasma downstream of the bow shock. Unlike the interplanetary solar wind, the magnetosheath plasma has a high $\beta$ and is relatively compressible. The MMS consists of four identical spacecraft, but here we do not utilize any multi-spacecraft technique. For each interval, we average over the values obtained from four spacecraft. We make use of the ion and electron plasma moments from the Fast Plasma Instrument~\citep{Pollock2016SSR}. We use ion velocity and temperature data to evaluate $\Mt$. But, whenever available, we use the electron number density to measure background density and density fluctuation. This is because the number density of electrons and ions are almost equal due to the quasi-neutrality of the plasma, and due to a larger thermal speed the electron density measurements are usually more accurate~\citep{Gershman2018PoP}. Magnetic field data are obtained from the fluxgate magnetometer (FGM)~\citep{Russell2016SSR, Torbert2016SSR} and combined with the plasma moments to obtain $\beta$ values. For example, we show the burst resolution MMS data (MMS1) obtained in the turbulent magnetosheath on 2017 January 18 from 00:45:53 to 00:47:43 UTC in \autoref{fig:mms}. The interval shows strong turbulent fluctuations with considerable density variations throughout the interval. The interval has $\beta \approx 19$.
       
    We eliminate intervals with very low average number density $n_e < 5$cm$^{-3}$, or very high average density $n_e > 50$cm$^{-3}$ because of larger uncertainties in such intervals. The length of every interval spans many times that of the typical correlation length (equivalently time) of the magnetosheath, which is about $\tau_c \sim 100$s~\citep[e.g.,][]{Stawarz2019ApJL}.

\subsection{Numerical Simulation}
    \paragraph{MHD plasma model}
    Along with the MMS observational data, we use a heavily modified version of the magnetohydrodynamical (MHD) code \textsc{flash} \citep{Fryxell2000,Dubey2008} that has recently been run on $10,\!080^3$ grid scales \citep{Beattie2024_10k}. Our code uses a highly-optimized, hybrid-precision \citep{Federrath2021}, positivity-preserving, second-order MUSCL-Hancock HLL5R Riemann scheme \citep{Bouchut2010,Waagan2011} to solve the compressible, ideal, MHD fluid equations in three dimensions,
    \begin{align}
    \partial_t \rho + \nabla\cdot\left(\rho \bfu\right) = 0,& \label{eq:continuity}\\
    \partial_t\!\left(\rho \bfu\right) + \nabla\cdot\left(\rho\bfu\!\otimes\!\bfu + p\mathbb{I} - \frac{1}{\mu_0}\bfb\!\otimes\!\bfb \right) = \rho \bm{ f},&\label{eq:momentum} \\
    \partial_t \bfb  + \nabla\cdot(\bfu\otimes\bfb - \bfb\otimes\bfu) = 0,& \label{eq:induction}\\ 
    \nabla\cdot\bfb = 0,& \label{eq:divb} \\
    p = \cs^2\rho+ \frac{1}{2\mu_0}\bfb \cdot \bfb,& \label{eq:pressure}
    \end{align}
    where $\rho$, $\bfu$, $\bfb$ and $\mu_0$ are the gas density, the velocity and magnetic fields, and the magnetic permittivity, respectively. Equation~\ref{eq:pressure} relates the scalar pressure $p$ to $\rho$ via the isothermal equation of state with constant sound speed $c_s$, as well as the pressure contribution from the magnetic field. We work in units $c_s = \rho_0 = \mu_0 = L = 1$, where $\rho_0$ is the mean gas density and $L$ is the characteristic length scale of the system, such that $L^3 = \mathcal{V} = 1$ is the volume. We discretize the equations over a triply periodic domain of $[-L/2, L/2]$ in each dimension, with grid resolution $10,\!080^3$ -- the largest grid in the world for simulations of this fluid turbulence regime. In order to drive turbulence, a turbulent forcing term $\bm{f}$ is applied in the momentum equation (details below). 
    
    Our numerical model is an implicit large eddy simulation (ILES), which relies upon the spatial discretisation to supply the numerical viscosity and resistivity as a fluid closure model. \citet{Shivakumar2023_numerical_dissipation} provide a detailed characterization of the numerical viscous and resistive properties of this solver by comparing the ILES model with direct numerical simulations (DNS), which have explicit viscous and resistive operators. They derived empirical models for transforming grid resolution. Using their relations, for our resolution, $N_{\rm grid}=10,\!080$, $\Re \in [1.4\times 10^6, 5.3\times 10^6]$ and $\Rm \in [1\times10^6, 4.5\times10^6]$.
    
\paragraph{Turbulent driving} 
    We drive the turbulence with a turbulent Mach number of $u_0/c_s = \Mt = 4.32 \pm 0.18 \approx 4$. This enables us to obtain scaling results in regimes where the simulation overlaps with both NI-MHD and weakly compressible MHD models, but also covers regimes in which the asymptotic approximations tend to break down. We apply a non-helical stochastic forcing term $\bm{f}$ in Equation~\ref{eq:momentum}, following an Ornstein-Uhlenbeck stochastic process \citep{Eswaran1988_forcing_numerical_scheme,Schmidt2009,Federrath2010_solendoidal_versus_compressive}, using the \textsc{TurbGen} turbulent forcing module \citep{Federrath2010_solendoidal_versus_compressive,Federrath2022_turbulence_driving_module}. The forcing is constructed in Fourier space such that kinetic energy is injected at the smallest wavenumbers, peaking at $\ell_0^{-1} = k_0L/2\pi = 2$ and tending to zero parabolically in the interval $1\leq kL/2\pi\leq3$. To replenish the large-scale compressible modes and shocks, we decompose $\bm{f}$ into its incompressible ($\nabla\cdot\bm{f}=0$) and compressible ($|\nabla\times\bm{f}|=0$) modes \citep{Federrath2010_solendoidal_versus_compressive}, and drive the turbulence with equal amounts of energy in each of the modes. 

    We set the correlation time of $\bm{f}$ to $t_0 = \ell_0/u_0$, such that the correlation time and turnover time of the largest eddy are equal. We drive the simulation into a statistically steady state, such that all of the first moments of underlying field variables no longer vary on average. We sample the turbulence 20 times across an $2t_0$ interval, averaging all statistics (e.g., $\Mt$ and $\delta\rho/\rho_0$ spectra, as discussed in \autoref{sec:results}) to ensure that our results are not sensitive to any intermittent events. We refer to \citet{Beattie2024_10k} for further information about the integral quantities and the time-evolution of the simulation.

\paragraph{Initial conditions \& steady-state magnetic field}\label{app:init_conditions}
    We initialize $\rho(x,y,z) = \rho_0$ and $\bfu = \bm{0}$. For our simulations, $B_0 = 0$, and only the turbulent $\delta\bfb$ remains. $\delta\bfb$ is maintained in a statistically stationary state via the turbulent dynamo \citep{Schekochihin2004_dynamo,Rincon2019_dynamo_theories,Kriel2022_kinematic_dynamo_scales}. The saturated state of our magnetic fields results in an Alfv\'en Mach number, $\Ma = u_0 / \Exp{v_A^{2}}{\V}^{1/2} = 2.03 \pm 0.04 \approx 2$, where $\Exp{v_A^{2}}{\V}^{1/2}$ is the rms Alfv\'en velocity. On volume-average, this provides $\beta \sim 1/8 \sim 1$. 

\paragraph{The inhomogeneous mass density field}
    \autoref{fig:dens} shows a two-dimensional slice of the logarithmic\footnote{Natural logarithmic being the most appropriate transformation of $\rho/\rho_0$, since we know that trans-to-supersonic turbulence gives rise to roughly lognormal distribution functions in mass density fluctuations \citep{Beattie2022_spdf}, with small correction to the higher-order moments based on the void statistics \citep{Hopkins2013_non_lognormal_s_pdf,Squire2017,Beattie2022_spdf}.}, mean-normalized mass-density field, $\ln(\rho/\rho_0)$, with in-plane magnetic field lines illustrated in white. The slice is taken from within the statistically stationary state of the turbulence. The strong inhomogeneities in the mass density can be observed from both the  numerous coherent structures (e.g., high-$\rho/\rho_0$ density filaments, shown in yellow and deep mass density voids, shown in green; both ubiquitous in supersonic MHD turbulence \citealt{Beattie2021_multishock}) presented in the simulation domain, and the roughly three orders of magnitude in $\rho/\rho_0$ that are resolved.
    
\section{Results}\label{sec:results}
\begin{figure}
	\centering
	\includegraphics[width = \linewidth]{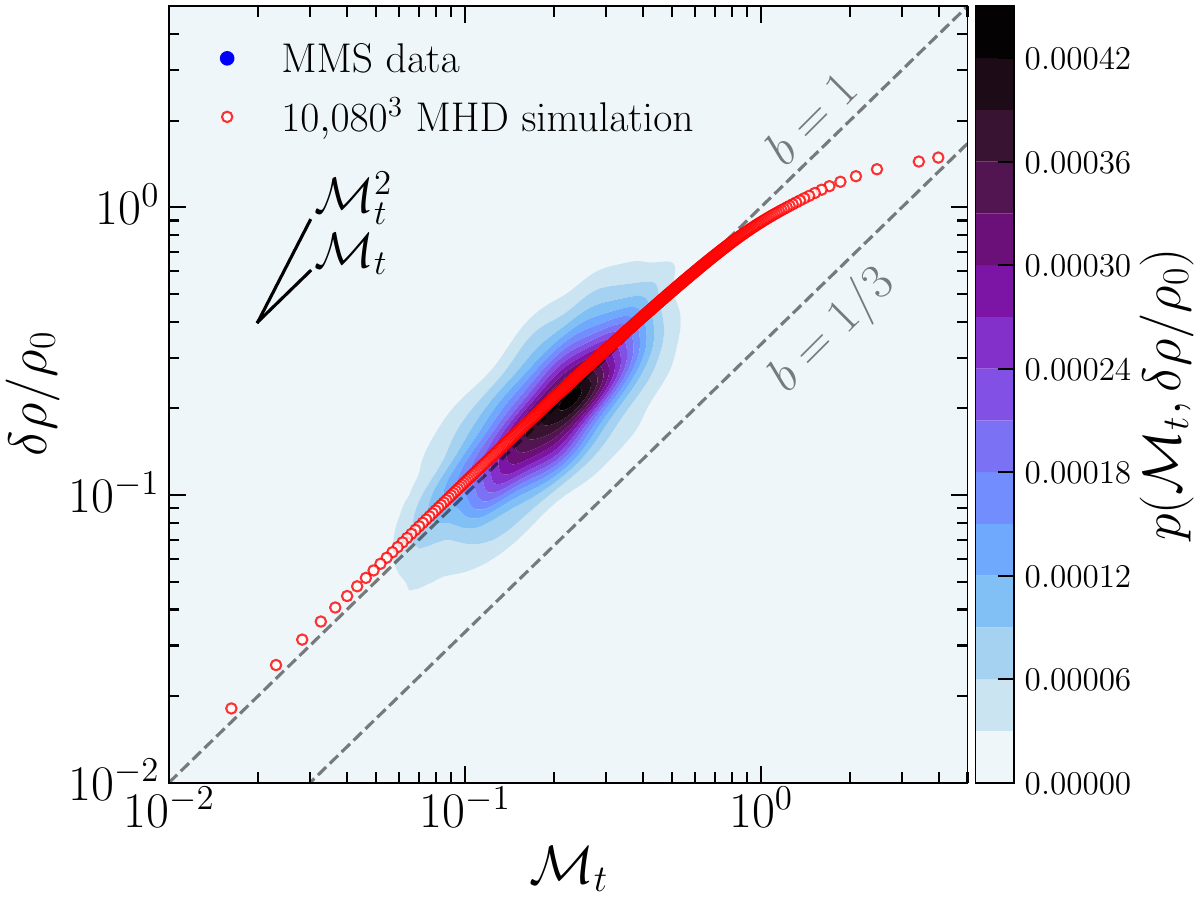}
	\caption{The variation of relative density fluctuations, $\delta\rho/\rho_0$, with turbulent Mach number, $\Mt$, in Earth's magnetosheath, measured by MMS (blue contours of the joint probability distribution function between, $\Mt$ and $\delta\rho/\rho_0$, $p(\Mt,\delta\rho/\rho_0)$) and from a $10,\!080^3$, highly-compressible MHD simulation (red dots; by combining \autoref{eq:scale-dependent-mach} and \autoref{eq:scale-dependent-density}). A best fit to the MMS data gives  $\delta \rho / \rho_0 = (0.83^{+0.54}_{-0.58}) \Mt^{0.92^{+ 0.30}_{-0.29}}$ and the simulation data yields $\delta\rho / \rho_0 =(0.97^{+0.23}_{-0.23})\Mt^{0.94^{+0.15}_{-0.15}}$. The two dashed lines of slope unity are shown for proportionality constants $b = 1/3$ and $b = 1$ (see \autoref{eq:linear_mach}).}
    \label{fig:scaling}
\end{figure}

    For each magnetosheath interval, we calculate the relative density fluctuation and turbulent Mach number, \autoref{eq:mach}. We approximately use 1200 magnetosheath intervals, creating a joint probability distribution function between $\delta \rho / \rho_0$ and $\Mt$, $p(\Mt,\delta \rho / \rho_0)$. In \autoref{fig:scaling} (blue) we show the contours of a Gaussian kernel density estimate for $p(\Mt,\delta \rho / \rho_0)$. The shade of the contours represents fixed values of the probability density for the underlying distribution. The contour plot qualitatively illustrates that the MMS data follow a linear relation closely between the two variables. We perform a maximum likelihood fit of the function $\delta\rho/\rho_0 = \theta_1 \Mt^{\theta_0}$ to the data and determine that the best fit parameters yield 
    \begin{align}\label{eq:mms_fit}
        \delta \rho / \rho_0 = (0.83^{+0.54}_{-0.58}) \Mt^{0.92^{+ 0.30}_{-0.29}},&& (\text{MMS data)}
    \end{align}
    consistent within $1\sigma$ to the $\delta \rho / \rho_0 \propto \Mt$ relation predicted by the weakly compressible MHD turbulence model \citep{Bhattacharjee_1998_weakly_compressible_solar_wind}, even in regimes where the asymptotic expansion in powers of the $\Mt$ of the turbulence may not be applicable. We show the corner plots for the fit in \autoref{app:fits}.

    For the simulation, we calculate the scale-dependent $\Mt$ and $\rho/\rho_0$ directly from the one-dimensional power spectrum of $\bfu$ and $\rho/\rho_0 - 1$. For example, the $\bfu$ spectrum is, 
    \begin{align}
        \mathcal{P}_{u}(k) = \int \d{\Omega_k}\; 4\pi k^2 \bfu(\bfk)\bfu^{\dagger}(\bfk),
    \end{align}
    where $\Omega_k$ is the integral over solid angle (see \citealt{Beattie2024_10k} for more information on $\mathcal{P}_{u}(k)$), such that the one-dimensional spectrum has a normalization, 
    \begin{align}
        \Exp{u^2}{\V} = \int\d{k}\;\mathcal{P}_{u}(k),
    \end{align}
    from Parseval's theorem. Hence, the scale-dependent $\Mt$ is simply,
    \begin{align}\label{eq:scale-dependent-mach}
        \Mt(\ell/L) = \frac{1}{c_s}\left(\int_{k}^{k_{u}}\d{k'}\;\mathcal{P}_{u}(k')\right)^{1/2},
    \end{align}  
    where $k_u$ is the microscale, or inner scale of the turbulence, which we define directly from the spectrum
    \begin{align}
        k_u = \left(\frac{1}{ \Exp{u^2}{\V}}\int \d{k}\;k^2 \mathcal{P}_{u}(k)\right)^{1/2},
    \end{align}
    such that the integral is not contaminated with modes in the viscous dissipation range. We similarly construct the scale-dependent rms for the mass density,
    \begin{align}\label{eq:scale-dependent-density}
        \frac{\delta\rho(\ell/L)}{\rho_0} = \left\langle\left(\frac{\rho(\ell/L)}{\rho_0}\right)^2\right\rangle = \left(\int_{k}^{k_{u}}\d{k'}\;\mathcal{P}_{\rho/\rho_0 - 1}(k')\right)^{1/2}.
    \end{align}  
    Plotting \autoref{eq:scale-dependent-density} as a function of \autoref{eq:scale-dependent-mach} allows us to directly compute the rms mass density as a function of $\Mt$ for each $\ell/L$ in the simulation. This process is equivalent to splitting the domain with volume $\V$ into sub-domains $\V=\bigcup_n\V_n$ and computing the rms for each collection of $\V_n$. Performing this analysis for all $\V_n$ means that instead of running multiple simulations with different plasma parameters (e.g., different $\M$, as in \citealt{Beattie2020,Beattie2022_spdf}), we can use our single highly-resolved simulation to study multiple combinations of $\delta\rho/\rho_0$ and $\Mt$. This is only possible with very high-resolution simulation data, with large amounts of dynamical range within the turbulence cascade, and allows us to test how robust $\delta\rho/\rho_0 \propto \Mt$ is, far away from the $k$ modes influenced by the driving mechanism.
    
    The computed $\delta \rho / \rho_0$ and $\Mt$ values are plotted with red, open markers in \autoref{fig:scaling}. Using the same maximum likelihood fitting process and model as in the MMS data, we find 
    \begin{align}\label{eq:sim_fit}
        \delta\rho/\rho_0 = (0.97^{+0.23}_{-0.23})\Mt^{0.94^{+0.15}_{-0.15}},&& (\text{simulation data)}
    \end{align}
    again showing that the simulation data is 1$\sigma$ consistent with the weakly compressible model, with a constant that is very close to unity. We discuss the proportionality constant further in \autoref{sec:disc}, and show the corner plots for the fits in \autoref{app:fits}.
    
    Both observation and simulation results shown here follow closely the scaling prediction from weakly compressible MHD theory with an inhomogeneous background field. Note that this is independent of the underlying $\beta$ ($\beta \sim 1$ for the simulation and $\beta \sim 10$ for the MMS data), suggesting that the prediction of the weakly compressible MHD theory is robust with respect to  $\beta$

\section{Discussion \& Conclusions}\label{sec:disc}
    Due to the weak $\delta\rho/\rho_0$, the interplanetary solar wind behaves very close to an incompressible fluid. Indeed, until the recent advent of Parker Solar Probe~\citep[PSP;][]{Fox2016SSR} data in the near-Sun solar wind, the relative amplitude of $\delta\rho/\rho_0$ has, for the majority of cases, remained much smaller than unity~\citep[e.g.,][]{Matthaeus1991JGR_NIMHD}. Therefore, weakly compressible MHD is usually adequate to describe the interplanetary solar wind fluctuations. However, in several astrophysical settings, where \textit{in-situ} measurements are not yet feasible, the compressive fluctuations can be significant, e.g.,  $\delta\rho/\rho_0$ varies by many orders of magnitude, as is the case in our simulations. Furthermore, $\beta$ is another parameter which is limited to a rather narrow band of values in the solar wind, but could be large in many astrophysical plasmas. Earth's magnetosheath provides an excellent natural laboratory to test high $\beta$, strongly compressive plasma environment using \textit{in-situ} data~\citep{Sahraoui2020RMPP_turbulence}.

    The proportionality constant, $b$, from \autoref{eq:linear_mach}, has been measured in a number of galactic \citep[e.g.,][]{Menon2020b} and even extra-galactic sources \citep[e.g.,][]{Sharda2021_driving_mode,Gerrard2023_driving_parameter_LMC}. In \autoref{fig:scaling} we show a very interesting effect. By measuring $\delta\rho/\rho_0 \propto \Mt$ across a broad range of scales in the turbulence (red markers) we find $\delta\rho/\rho_0 \approx 1/3\Mt$ on the outer scale ($b=1/3)$, indicating incompressible driving, and then it changes to $\delta\rho/\rho_0 \approx \Mt$ ($b=1)$ on smaller scales (smaller $\Mt$ and $\delta\rho/\rho_0$). This value of $b$ is theorized to be a signature of when the driving mechanism ($\bm{f}$ in \autoref{eq:momentum}) for the turbulent fluctuations is compressive. This means that deep within the cascade $\delta\rho/\rho_0 = \Mt$, regardless of the nature of $\bm{f}$ (a mix of compressible and incompressible modes in our simulation). Based on the compressible / incompressible mode decomposition for the simulation in \citet{Beattie2024_10k}, on the scales that $\delta\rho/\rho_0 = \Mt$ the turbulence is dominated by incompressible modes, so this demonstrates that even if the momentum modes are mostly incompressible, if one measures $b$ deep in the cascade, where there is no driving source, one gets the compressible-driving relation $\delta\rho/\rho_0 = \Mt$.

    In this paper, we investigate the scaling of $\delta\rho/\rho_0$ with the turbulent Mach number in high $\beta$, highly-compressible regime. This has been accomplished using a large number of MMS measurements (over 1200 intervals) in Earth's magnetosheath and an unprecedentedly high-resolution $10,\!080^3$, highly-compressible MHD simulation \citep{Beattie2024_10k}. Not only do both the \textit{in-situ} data as well as the simulation support the scaling $\delta \rho / \rho_0 \propto \Mt$ predicted by the weakly incompressible theory in the presence of background inhomogeneities~\citep{Bhattacharjee_1998_weakly_compressible_solar_wind}, but they also agree within 1$\sigma$ with one another, even in proportionality constants. This might be due to the fact that both the magnetosheath plasma and the simulation used here possess abundant inhomogeneities, e.g., mass density filaments, voids, current sheets, sheared flows, and other coherent structures. The structures and strong variations in the plasma variables render the plasma sufficiently inhomogeneous such that linear $\sim \Mt$ scaling becomes valid instead of quadratic $\sim \Mt^2$ scaling, as we show in \autoref{fig:scaling}. This demonstrates both the robustness of the \citet{Bhattacharjee_1998_weakly_compressible_solar_wind} model for a variety of $\beta$ plasmas, and the broad applicability of the $10,\!080^3$ MHD dataset to both space and astrophysical plasmas.

\section*{\textbf{Acknowledgments}}\label{sec:ack}
    The data used in this analysis are Level 2 FIELDS and FPI data products, in cooperation with the instrument teams and in accordance with their guidelines. All MMS data are available at \href{https://lasp.colorado.edu/mms/sdc/}{https://lasp.colorado.edu/mms/sdc/}. This research was supported in part by the NASA Heliospheric GI Grant No. 80NSSC21K0739 and NASA Grant No. 80NSSC21K1458.
    J.~R.~B. acknowledges the high-performance computing resources provided by the Leibniz Rechenzentrum and the Gauss Center for Supercomputing grant~pn76gi~pr73fi and pn76ga, and Compute Ontario and the Digital Research Alliance of Canada (alliancecan.ca) compute allocation rrg-ripperda. J.~.R.~B. and A.~B further acknowledge the support from NSF Award 2206756. 

\software{Data analysis and visualization software used in this study: \textsc{C++} \citep{Stroustrup2013}, \textsc{numpy} \citep{Oliphant2006,numpy2020}, \textsc{numba}, \citep{numba:2015}, \textsc{matplotlib} \citep{Hunter2007}, \textsc{cython} \citep{Behnel2011}, \textsc{visit} \citep{Childs2012}, \textsc{scipy} \citep{Virtanen2020},
    \textsc{scikit-image} \citep{vanderWalts2014}, \textsc{cmasher} \citep{Velden2020_cmasher}, \textsc{pandas} \citep{pandas2020}, \textsc{joblib} \citep{joblib}, \textsc{emcee} \citep{Foreman-Mackey2013_emcee}, \textsc{corner}\citep{corner}}


\appendix
\section{Maximum Likelihood Fits}\label{app:fits}

\begin{figure}
        \centering
        \includegraphics[width=0.9\linewidth]{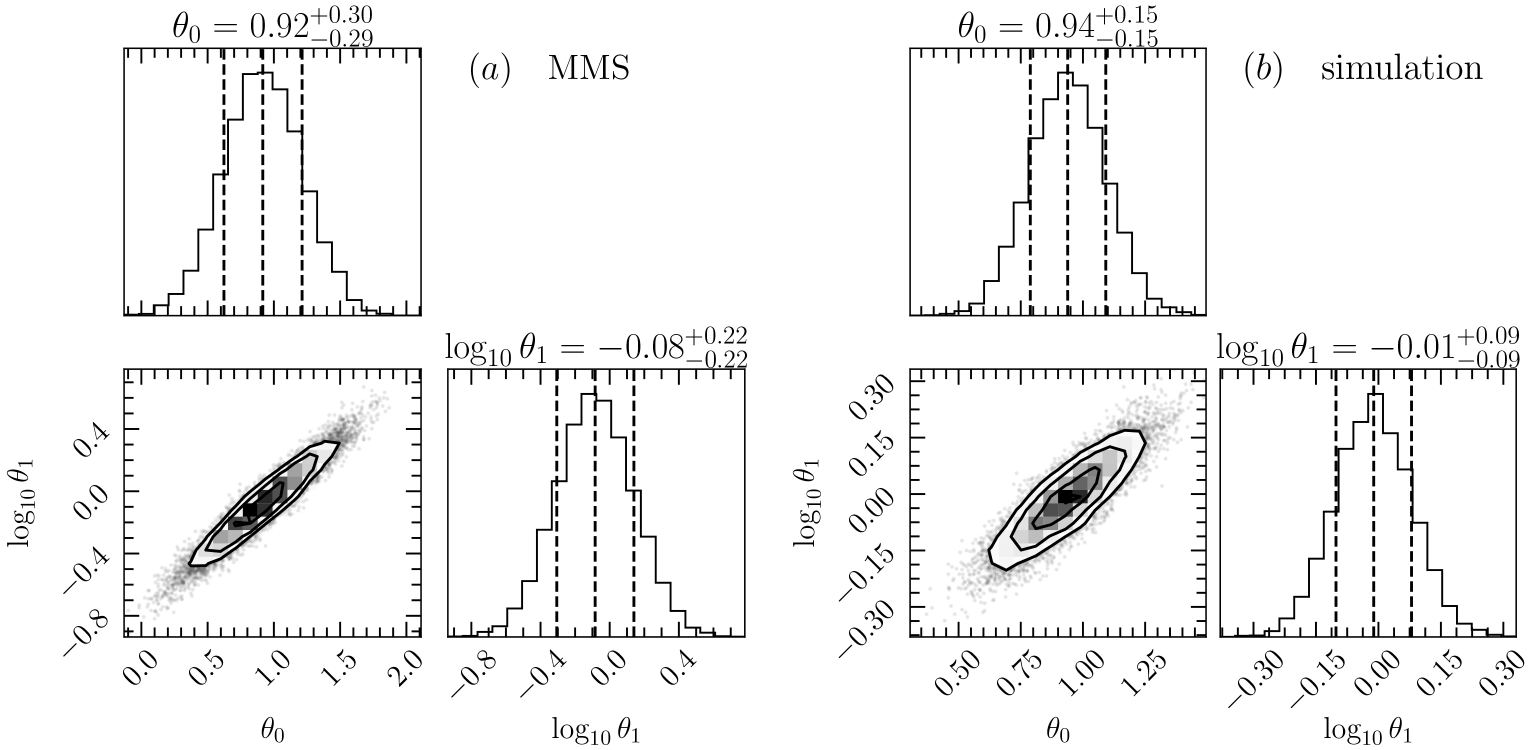}
        \caption{The corner plots for the two $\log_{10} \delta\rho/\rho_0 = \theta_0 \log_{10}\Mt + \log_{10}\theta_1$ fits (shown in \autoref{eq:mms_fit} and \autoref{eq:sim_fit}) to the MMS (a) and simulation data (b), showing that (1) within 1$\sigma$, $\delta\rho/\rho_0 = \Mt$, as predicted from weakly compressible turbulence theory \citep{Bhattacharjee_1998_weakly_compressible_solar_wind}, and (2) that with 1$\sigma$ both the MMS and simulation data agree with one another.}
        \label{fig:corners}
    \end{figure}

    We fit the simple model $\delta\rho/\rho_0 = \theta_1 \Mt^{\theta_0} \iff \log_{10} \delta\rho/\rho_0 = \theta_0 \log_{10}\Mt + \log_{10}\theta_1$ to both the MMS and simulation data, to primarily test if there is statistical agreement between the data and the weakly incompressible theory from \citet{Bhattacharjee_1998_weakly_compressible_solar_wind}. Naturally, we are able to also probe the similarities and differences between the MMS data and the simulation, too. We use a maximum likelihood approach, utilizing the \citet{Foreman-Mackey2013_emcee} sampler. We have very weakly constrained priors, with $-4 \leq \theta_0 \leq 4$ and  $-2 \leq \log_{10}\theta_1 \leq 1$, ensuring that we do not over-constrain the fits. We use a regular likelihood function, and we sample the posterior with 32 walkers, for 5000 steps, getting rid of the first 500 steps as a burn-in stage. We show the posterior for the MMS (a) and simulation (b) panels in \autoref{fig:corners}, which indicates that the posterior is well-sampled, not over-constrained, and the parameters have a strong covariance (as expected). As we discussed in the main text, the fits show two key aspects. Firstly, that (within 1$\sigma$) $\delta\rho/\rho_0 = \Mt$, as predicted from weakly compressible MHD theory \citep{Bhattacharjee_1998_weakly_compressible_solar_wind}, and secondly, both the MMS and simulation data agree with one another.

\end{document}